# Modelling crystallization: When the normal growth velocity depends on the supersaturation

V. V. Ivanov[1], C. Tielemann[2,3], K. Avramova[5], S. Reinsch[4], V. Tonchev[1,*]

[1]Faculty of Physics, Sofia University, 1164 Sofia, Bulgaria

[2]Clausthal University of Technology, Institute of Non-Metallic Materials, 38678 Clausthal-Zellerfeld, [3]Advanced Semiconductor Materials Lithography, 12347 Berlin, Germany

[4] Bundesanstalt für Materialforschung und prüfung (BAM), 12205 Berlin, Germany

[5]Institute of Physical Chemistry, Bulgarian Academy of Sciences, 1113 Sofia, Bulgaria

* corresponding author: tonchev@phys.uni-sofia.bg

**Abstract.** The crystallization proceeds by the advance of the crystal faces into the disordered phase at the expense of the supersaturation. Using a conservation constraint for the transformation ratio $\alpha$ as complementing the rescaled supersaturation to 1 and a kinetic law for the normal growth velocity as function of the supersaturation raised to power $g$ – the growth order, we derive a general equation for the rate of transformation $d\alpha/dt$. We integrate it for the six combinations of the three spatial dimensions $D$ = 1, 2, 3, and the two canonical values of the $g$=1, 2, towards obtaining expressions for $\alpha_{Dg}$. The same equation, with $g$=1 and $D = n$, the so called Avrami exponent, is obtained when taking only the linear in $\alpha$ term from the Taylor's expansion around $\alpha$=0 of the model equation of Johnson-Mehl-Avrami-Kolmogorov (JMAK). We verify our model by fitting datasets of $\alpha_{21}$ and $\alpha_{31}$ (from 0 to $\alpha_{upper}$=0.999) with JMAK to obtain from the fit $n$ = 1.725, 2.43, resp. We show further how the values of $n$ depend on the value of $\alpha_{upper}$ to which the fit is performed starting always from 0. Towards building a validation protocol, we start with validating $\alpha_{21}$ with published results.

**Introduction**

Crystallization is a process of first order phase transformation from a disordered to an ordered, crystal phase [1]. One of its aspects is the significant symmetry reduction– from the highest for the embedding space, circular ($O_2$) or spherical ($O_3$) - that of the disordered state, down to the symmetry group of the forming crystal. It is the difference in the chemical potentials of the two phases that drives the transformation upon leveling the chemical potential when the equilibrium is re-established. At the end of the 20's a key concept in the theory of crystal growth was formulated by Walter Kossel [2], and, independently, by Ivan N. Stranski [3] – they identified the so called *half-crystal* or *kink* crystal position as the gear of the reproducible crystal growth – the attachment of a crystal building unit, atom or molecule, in/to the kink position creates the same number of bonds that the unit has in the volume of the crystal, since there, in the volume of the crystal, each bond is shared between the two units on the each side of the bond. Instead of adding an illustration of what is kink we can imagine the crystal forming unit in the volume of the crystal and the cut one bond of any bond pair in given direction in such a way that the remaining bonds form a concavity for this atom ("half-volume, half-plane, half-step"). It is different from the so called nucleus (seed) since the kinks perpetuate the growth started from the nucleus. Of course, somewhere there, between the nucleus and the crystal is the first kink but with respect to the present text this is a rather philosophical problem.

As a result, this lowers the energy of the attaching unit, releasing in the ambience the so called *latent heat* (equal to the energy of the created bonds). Turning the focus towards the disordered phase, such an attachment-to-kink (A2K) event leaves behind a kink again if not reaching the end of the crystal. Through such a sequence of elementary acts the crystal symmetry is reproduced along the increasing length-scale of the phenomenon, from atomic to macroscopic. It is the introduction of the concept of kink that paved the way to build the new, *molecular-kinetic* theory of crystal growth, where the kinks play a key role [4,5] within the terrace-step-kink [6] paradigm that eventually led to the seminal model of Burton, Cabrera and Frank (BCF) [7,8]. Note that the "BCF thinking" is embedded in (1+1)D and, therefore, it does not distinguish on the conceptual level between step and kink.

As already mentioned, crystals are finite objects and the *crystal growth events* – the attachments to kink positions, are not enough for the process to be a self-sustaining one. Other, complementary events of attachment to positions where the attaching unit is less bound to the crystal than in the kink positions are needed. They could be united under the general name of *kink generation* events, regardless of what they are called in the varying contexts – *aggregation*, *nucleation*, etc. The process of kink generation could be rather complex one and require multiple stages, especially during the growth of three-dimensional crystals since the attachment of a single crystallizing unit to a smooth on the atomic scale crystal plane does not result in a kink. This is why the various realizations of the so called epitaxy employ as substrates ordered arrays of steps (and kinks on them) – the vicinal crystal surfaces, see for further discussion [9,10] and the references therein.

Yet the first joint paper by R. Kaischew and I. N. Stranski [11] provides a criterion that could be used, in fact, to distinguish between the crystal growth (crystallization) and the aggregation. The original formulation defines the units which are bound to the crystal with fewer bonds than in a kink position as not belonging to the *equilibrium crystal shape* [6]. In the thought experiment proposed in [11] of removing such units one should also remove these that leave behind units less bound than in a kink position. Here a simple argument is provided in order to illustrate the approach of [11]: if one removes the units described above from the crystal until ending up with a flat crystal plane or a vicinal crystal surface with only kinks on the steps further removals are impossible and vice versa - since no planes can be observed at any stage of the aggregate growth, even if the growth is simulated on a regular lattice [12,13], the aggregate will be destroyed completely by such a thought (or translated *in silico*) experiment. From the same point of view, in 1D it is impossible to distinguish in between crystallization and aggregation, and, in fact, from nucleation – everything that can happen in 1D is attachment to one of the two ends of the growing 1D "rod".

In the 30's started a parallel theoretical development to meet the needs of the physical metallurgy in the description of phase transformations, including re-crystallization ("order out of order"). The early development and success is mainly due to Johnson, Mehl, Avrami and Kolmogorov (JMAK) [14–16] and nowadays the number of papers that describe what the original modelling was designed for and how the consecutive applications went well beyond the initial prerequisites, as for example the constant interfacial velocity, recrystallization, is steadily increasing, see a general discussion of JMAK in [17–20]. When started the use of the model for modelling the nucleation kinetics, the so called "N-t curves" – the dependence of the number of nuclei $N$ of the time $t$, is beyond the scope of the present study but could provide an interesting perspective on the model itself.

Here we start with the classical formula [21–25] of JMAK:

$$\alpha = 1 - \exp\left[-kt^n\right] \tag{1}$$

with $n$ = 1, 2, 3 and 4 being the dimensionality of the space embedding the process plus 0/1 to account for the absence/presence of a parallel nucleation. As pointed out by R. Svoboda in [21] this expression is based on the concept of the extended volume, see for further details on this the review of M. Fanfoni and M. Tomellini [19]. We add to the abbreviation of the model the (stretched or Avrami) exponent n to which the time is raised, to arrive at the convenient form of abbreviation JMAKn.

This expression acquired an enormous spread in the years providing [22] a flexible sigmoid curve with two fitting parameters, or three if subtracting from the time $t$ some initial (induction) period $t_0$, or even four if presenting $\alpha$ as $\alpha \equiv N(t)/N_{max}$ as in the case of nucleation. Especially the relaxation of the restrictions on $n$ – from integer-valued to real-valued one, leads to reporting various non-integer values with a "side effect" - the weird dimension of $k$ that is expected to encode the dependence on the temperature - $[k] = \text{Time}^{-n}$, that is why sometimes this dimension is omitted [26] but what physical meaning a kinetic constant with the dimension of Time$^{-1.725}$ would have? The direct solution is to leave $k$ in the inmost parentheses – $(kt)^n$ as done, for example, in [27].

Among the other alternatives of modelling growth phenomena, it is worth mentioning here the general purpose three-parameter model of Richards [12] with special cases – the Verhulst [28] and Gompertz [29] models. It is also a four-parameter one if it is necessary to include also the maximal value of the quantity that is modelled, especially when modelling populations and their dynamics.

In 2013 Nanev and coworkers [24, 25] used, as part of their protocol for growing insulin crystals of certain size from solutions ("order out of disorder"), a model for the growth and dissolution of *N* equally-sized crystals in 3D, derived with the assumption that the supersaturation is not sustained, i.e. it is raised to a maximal value in the beginning of the crystallization. The prerequisites behind the model are simple – to formulate the mass balance by expressing the current concentration as function of the size of the crystals already grown and then, to plug it in into the expression for the growth rate based on the kinetic law for the normal to the crystal face velocity. In this way was obtained [30] a differential equation for the rescaled crystal size $L \equiv l/l_{max}$ that is solved in the simplest case – for growth order *g* = 1, obtaining an expression for the dimensionless time $T(L)$. Unfortunately, in 3D it is technically impossible to go further with obtaining analytical expression for $L(T)$ [32]. Additionally, in [30] a dependence was also obtained that links the supersaturation to the crystal size and there it is was pointed out that such a dependence could serve to follow the evolution of the crystal size by monitoring the supersaturation.

A model in 1D with same prerequisites was employed by Kashchiev [33] and, as a result, he obtained JMAK1. Another comment is due here concerning the need to deal with care when modeling in this way the crystallization in reduced dimensions, *D* = 1, 2. It is of particular importance to have the diffusion field in the same constrained dimensionality otherwise the applicability of the kinetics law may be questionable.

**I. The model**

What distinguishes the first order phase transition from the second order one is the coexistence of domains of both phases at the transition point when the transition is of first order while when it is of second order the two phases become indistinguishable there. Driving the system away from the transition point makes one of the two phases stable and the other one – metastable (when the deviation is still *small*) or unstable. Then the domains of the stable phase grow invading the metastable one and it is the difference in the chemical potentials $\Delta \mu = \mu_m - \mu_s$ (*s* stands for the stable and *m* for the metastable phase) of the two phases that quantifies the *driving force* of the process. Crystallization is archetypical case of a first order phase transition. On atomic level and despite the different realizations, the crystal building units have to attach to the crystal, mainly to the kink positions from the crystal faces. The alternative route is in the course of *kink generation* events that follow the inevitable disappearance of the kinks at the ends of the finite crystals - attachment to positions in which the units are less bound than in a kink. Thus the crystallizing particles/units leave the crystal surrounding, i.e. the disordered phase. As a result and when the supersaturation is not controlled, their number decreases there. That is why it is widely accepted to use the so called (relative) supersaturation [34] $\sigma = (C - C_e)/C_e$ as a measure of the distance to equilibrium - when the concentration acquires its equilibrium value $C = C_e$ thus the growth seizes, and apart from the discussion of applicability of concentration (and not other quantities such as the activity, solubility, etc). Same type of an expression for the supersaturation is widely used to find the step velocity in the various sophistications of the models of step flow growth [35] but there the equilibrium or reference [36] concentration $C_e$ could account for the effect of the step-step interactions.

The starting point of our considerations is the expression [30] for the normal growth velocity - the velocity of advancement of the crystal face(s) into the disordered ambience, as function of the supersaturation σ:

$$r = \beta \sigma^g \qquad (2)$$

where $\beta$ is a proportionality (kinetic) constant with the dimension of r, $[r]$ = LT$^{-1}$, $g$ is the so called *growth order*. Note, that in β is comprised a coefficient that accounts for the density change [30,35] during the phase transformation, usually called *molecular volume*. The canonical values of $g$ from Burton, Cabrera and Frank [7] are 1 in the regime of instantaneous kinetics (the diffusion limited regime of the growth) and 2 in the kinetically, or attachment-detachment limited one. Due to the resemblance to the way the kinetics of the chemical reactions is described in the law of mass action of Guldberg and Waage [30], the power $g$ is called growth order although, formally, on the left hand side of eq.(2) is the velocity of the front of a first order phase transition and not the time derivative of a concentration as in the context of the chemical kinetics. Still, this motion is only an effective one in the sense that each unit remains fixed after its attachment to the crystal thus mediating on the atomic scale the phase transformation, hence the motion of the crystal face.

Our strategy will be first to assume that the number of growing non-interacting through the diffusion field centers is fixed and then to obtain a general equation for the time evolution of the transformation ratio $\alpha \in [0,1]$, see eq. (16) below, and we will use a mass conservation relation to exclude the rescaled supersaturation $\sigma/\sigma_0 \in [0,1]$ from the considerations that follow:

$$\alpha = 1 - \sigma/\sigma_0 \tag{3}$$

This permits to continue with expressing for the normal growth velocity $r$ via the rescaled supersaturation:

$$r/r_0 = (\sigma/\sigma_0)^g \tag{4}$$

and, therefore:

$$r = r_0 (1-\alpha)^g \tag{5}$$

The overall growth rate $G$ defined in terms of time derivative of the characteristic crystal size $l$ is:

$$G \equiv dl/dt \tag{6}$$

and could be obtained as twice the normal growth velocity assuming that any two crystal faces remain parallel to each other in the course of the so called *polyhedral* growth during which the crystals grow in each of the two opposite directions preserving their polygonised shapes:

$$G = 2r_0 (1-\alpha)^g \tag{7}$$

In experiments when the supersaturation is driven to its maximal value in the beginning of the growth and not sustained further as in the so called *batch crystallization* mode [37], and the parallel nucleation in the volume is suppressed, as in the variants of the so called *double impulse technique* [38], a fixed amount of $N$ crystals is growing in parallel retaining the same characteristic size $l$ [30]. One of the conditions for this to be fulfilled is $N$ to be small in order to not allow for the overlapping of the concentration fields around each of the growing crystals [30] and while the diffusion is slow (compared to the growth kinetics), hence $g = 1$, this is more likely to happen. For example, when the diffusion of the incorporating units is slow *a priori*, as in the case of protein crystallization, $g=1$ is preserved along almost the whole range of studied supersaturations, see [39] (after linearization of the axes of their figure 4, otherwise the authors obtained $g=3$ in log-log coordinates fitting through the whole range of points). Thus, an additional constraint in our model is on the value of $g$ - it remains fixed throughout the whole process. Still, one should have in mind that $g$ could change with the decrease of the supersaturation [7] but this is a subject of a parallel study. The growth will seize when the system is again at equilibrium and the maximal value of the characteristic crystal size $l_{max}$ is achieved [30]:

$$l_{max} = \left(V_m \frac{C_0 - C_e}{N}\right)^{1/D} \tag{8}$$

Note that the maximal crystal size $l_{max}$ is a function of the initial concentration difference [30] - a material excess that will be shared among the $N$ identical and independent copies of the same growing crystal. This time the molecular volume $V_m$ is written explicitly.

We proceed now with non-dimensionalization [40] by introducing a dimensionless quantity – the rescaled size $l/l_{max}$ and dividing and multiplying the right hand side of equation (6) with $l_{max}$ to combine it with (7):

$$\frac{l_{max} d(l/l_{max})}{dt} = 2r_0 (1-\alpha)^g \tag{9}$$

Thus, from (9) arises in a natural way the time scale of the phenomenon:

$$\tau_{Dg} \equiv \frac{l_{max}}{r_0(g)} = \left(V_m \frac{C_0 - C_e}{N}\right)^{1/D} \Big/ \beta\left[(C_0 - C_e)/C_e\right]^g \tag{10}$$

a composite parameter comprising through $r_0$ and $l_{max}$ the initial concentration excess, the number of growing crystals $N$, and including also the growth order $g$ and the kinetic coefficient $\beta$:

$$\tau_{Dg} \equiv \frac{l_{max}}{r_0(g)} = \frac{V_m^{1/D}}{C_e^g \beta} \left(\frac{1}{N}\right)^{1/D} (C_0 - C_e)^{\frac{1}{D}-g} \tag{11}$$

It is important to stress that the dimensions of quantities used in eq. (10) should be tweaked with regards to their dimensions in order to arrive at the dimension of length in the numerator. This could be achieved if all three quantities in the parentheses will be defined with the dimension of Length$^{-D}$. In order to compare the time scale (10) with the time scale of the JMAKn model from [41] we provide here its formula as defined for the case of growth of a constant number of crystals *N*:

$$\tau_{JMAKn} \sim \frac{1}{UN^{1/D}} \tag{12}$$

where *U* is the constant interfacial velocity – an intrinsic property of the JMAK model. Note also that the two time scales, (10) and (12), differ overall by a factor of ~1.1 as will be seen below.

Now the differential equation (9) is written in a non-dimensional form (omitting the indices of $\tau$ for simplicity):

$$\frac{d(l/l_{max})}{d(t/\tau)} = 2(1-\alpha)^g \tag{13}$$

On the other hand, $\alpha$ is simply the rescaled volume of the crystal phase $\alpha = (l/l_{max})^D$ [42] provided that the number of the growing crystals *N* is fixed from the beginning of the crystallization $l/l_{max} = \alpha^{1/D}$:

$$\frac{d(\alpha^{1/D})}{d(t/\tau)} = 2(1-\alpha)^g \tag{14}$$

and, performing the differentiation in the numerator of the left hand side of (14), to arrive at:

$$\frac{\alpha^{(1-D)/D} d\alpha}{D d(t/\tau)} = 2(1-\alpha)^g \tag{15}$$

Thus we obtain finally:

$$\frac{d\alpha}{d(t/\tau)} = 2D\alpha^{(D-1)/D}(1-\alpha)^g \tag{16}$$

One can modify further (16) by adding indices to $\alpha$ and $\tau$ in order to denote them as *D*- and *g*-specific:

$$d\alpha_{Dg}/d(t/\tau_{Dg}) = 2D\alpha_{Dg}^{(D-1)/D}(1-\alpha_{Dg})^g \qquad (17)$$

Equation (16) comprises the combined action of two opposite feedback mechanisms – a positive (auto-catalytic) and a negative (self-limiting) one. The positive feedback term is $2D\alpha_{Dg}^{(D-1)/D}$ independently of the value of *g*. In *D*=1 it is 2, so there is no positive feedback and the one-dimensional "crystal" grows via consecutive attachments in the two endpoints at any stage of the process. In *D*=2 the positive feedback is $4\alpha_{2g}^{1/2} = 4(l/l_{max})$ and this is the rescaled perimeter of the growing crystal(s), it increases during the growth and thus the rate of transformation increases as well. In *D*=3 it is $6\alpha_{3g}^{2/3} = 6(l/l_{max})^2$ - the rescaled area of the 6 squares enclosing the growing cube. For *D*=2, 3 and $g=1$ the negative feedback results from the multiplication of the positive one with $\alpha_{D1}$ taken with a negative sign. For $g=2$ and $\alpha$ still close to 0, the negative feedback is already $2\alpha$-times the positive feedback and this is why the $g=2$ curves for each *D*=2, 3 are having their maximal values for lower values of $\alpha$ compared to the $g=1$ curves, and their magnitudes are also lower, see Figure 1. When $\alpha$ increases approaching 1, an additional term, proportional to $\alpha^3$, adds to the positive feedback and this explains why there are observed inflection points on the two curves meaning slowing the decrease of the magnitude due to this additional term. This effect may look like a change in the growth mechanism but it is not.

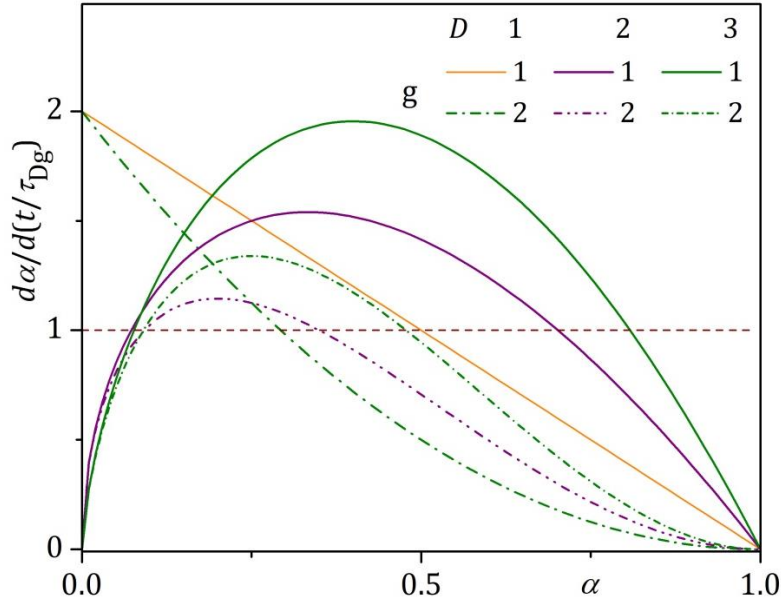

Figure 1. The phase space of our model - six different time derivatives of $\alpha$, following from equation (17). Also shown is the constant rate, independent of $\alpha$, it corresponds to the line $\alpha = t/\tau$ on the integral plot, Figure 2.

Towards the integration of (17) we separate the variables:

$$\frac{d\alpha_{Dg}}{\alpha_{Dg}^{(D-1)/D}\left(1-\alpha_{Dg}\right)^{g}} = 2Dd\left(t/\tau_{Dg}\right) \tag{18}$$

A general solution of the differential equation (18) can be found in the following form:

$$t(\alpha) = \frac{1}{2D}\mathrm{B}\left(\alpha;\frac{1}{D},1-g\right) \tag{19}$$

Here $\mathrm{B}(x;a,b) = \int_0^x t^{a-1}(1-t)^{b-1}dt$ is the incomplete Euler beta function. Solving for $\alpha(t)$ we obtain:

$$\alpha(t) = \mathrm{B}^{-1}\left(2Dt;\frac{1}{D},1-g\right) \tag{20}$$

where B$^{-1}$ is the inverse of the beta function. This solution is general but a closed-form expression usually cannot be obtained for an arbitrary choice of $D$ and $g$. Instead of this we develop a numerical procedure [43] based on the differential form, eq. (16), to produce $\alpha_{Dg}$ for real-valued $D$'s and $g$'s. Analytically it is still possible to integrate (18) for the three integer, physically justified, values of $D$ = 1, 2 and 3, combined with one of the two canonical values of $g$ = 1 or 2, see the results in Table 1. Note that for $D+g \leq 3$ expressions for $\alpha(t)$ are presented there while for $D+g > 3$ only expressions for $t(\alpha)$ are obtained. Note further that $\alpha_{11}$ coincides, in fact, with JMAK1 and we will preserve this similarity when working with JMAKn. We will show that with the increase of $D$ (above 1), the divergence between the two models increases, Table 4.

Table 1. Integral behavior of the model for the six combinations of the spatial dimension $D$= 1, 2, 3 and $g$ - the growth order with canonical values 1 and 2 [7]. Closed-form expressions for $\alpha(t)$ are obtained only for $D + g \leq 3$, the three shaded in grey cells, while for the other three cases only expressions for $t(\alpha)$ are achieved by the integration.

| $D$ | $g$ | |
|---|---|---|
| | 1 | 2 |
| 1 | $\alpha_{11} = 1 - \exp(-2t/\tau_{11})$ | $\alpha_{12} = \dfrac{2t/\tau_{12}}{2t/\tau_{12}+1}$ |
| 2 | $\alpha_{21} = \tanh^2(2t/\tau_{21})$ | $t/\tau_{22} = \dfrac{1}{4}\left(\dfrac{\alpha_{22}^{1/2}}{(1-\alpha_{22})} + \tanh^{-1}\alpha_{22}^{1/2}\right)$ |
| 3 | $\dfrac{t}{\tau_{31}} = \dfrac{1}{12}\left[\ln\left(\dfrac{\alpha_{31}^{2/3}+\alpha_{31}^{1/3}+1}{(1-\alpha_{31}^{1/3})^2}\right) + 2\sqrt{3}\tan^{-1}\left(\dfrac{\sqrt{3}\alpha_{31}^{1/3}}{2+\alpha_{31}^{1/3}}\right)\right]$ | $\dfrac{t}{\tau_{32}} = \dfrac{1}{18}\left[\dfrac{3\alpha_{32}^{1/3}}{1-\alpha_{32}} + \ln\left(\dfrac{\alpha_{32}^{2/3}+\alpha_{32}^{1/3}+1}{(1-\alpha_{32}^{1/3})^2}\right) + 2\sqrt{3}\tan^{-1}\left(\dfrac{\sqrt{3}\alpha_{32}^{1/3}}{2+\alpha_{32}^{1/3}}\right)\right]$ |

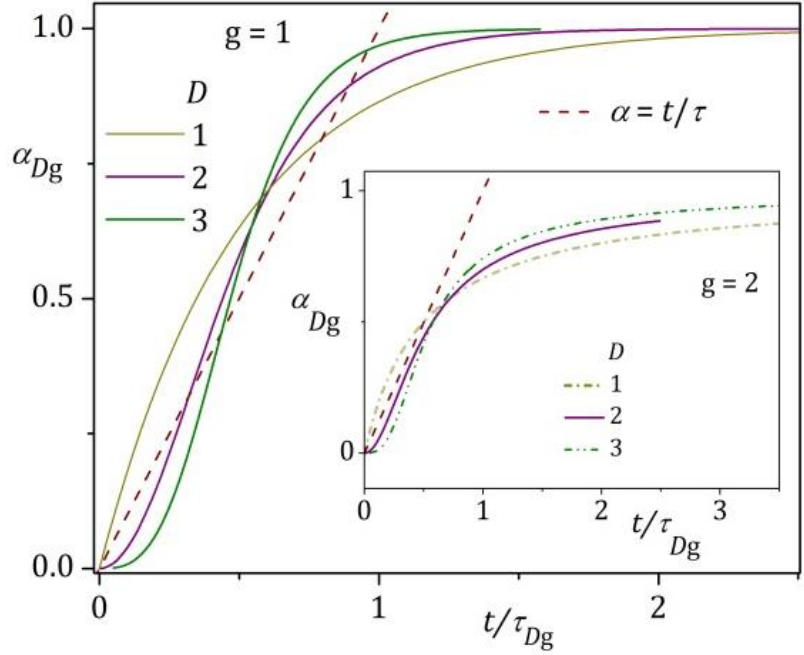

Figure 2. The integral curves $\alpha_{Dg}$ of our model. Main figure: g =1, inset: g = 2. The line $\alpha = t/\tau$ is not only a "guide to the eye" – here, for example, it serves as a reference line to compare the behavior of the curves for the different values of *g*.

In order to obtain the inflection points of the model curves above, Figure 2, we differentiate both sides of equation (18) to obtain:

$$\alpha'' \equiv d^2\alpha/dt^2 = -2\alpha^{-1/D}(1-\alpha)^{g-1}\left[(Dg+D-1)\alpha - D+1\right]\alpha' \qquad (21)$$

Substituting the first derivative from and simplifying further:

$$\alpha'' = -4D\alpha^{\frac{D-2}{D}}(1-\alpha)^{2g-1}\left[(Dg+D-1)\alpha - D+1\right] \qquad (22)$$

A necessary but not sufficient condition for an inflection point $t^*$ to exist is:

$$\alpha''^*(t^*) = 0$$

Therefore, we obtain the equation:

$$-4D\alpha^{*\frac{D-2}{D}}(1-\alpha^*)^{2g-1}\left[(Dg+D-1)\alpha^* - D+1\right] = 0 \qquad (23)$$

Solving it for $\alpha^*$ analytically gives the three possible solutions below, which we will analyse:

$$\alpha = 0, \text{ if } D > 2$$

$$\alpha = 1, \text{ if } g > 1/2$$

$$\alpha = \frac{D-1}{Dg + D - 1} \quad (24)$$

$\alpha = 0, 1$ cannot be the values at the inflection point and the only possible candidate left is eq.(24).

It should be noted here that if $D = 1$ for any value of $g$, $\alpha^* = 0$ and, therefore, all $\alpha_{1g}$ models do not have an inflection point.

From above we have a general solution for $t = f(\alpha, g, D)$. Substituting (24) in (19) we obtain that if an inflection point exists, it will have the following coordinates:

$$\{t^* | \alpha^*\} = \left\{ \frac{1}{2D} B\left( \frac{D-1}{Dg+D-1}; \frac{1}{D}, 1-g \right) \Big| \frac{D-1}{Dg+D-1} \right\} \quad (25)$$

Using further this notation we show the values of the $t^*$ and $\alpha^*$, see Table 2 and Figure 3 below.

Table 2. Inflection points of the $\alpha_{Dg}$ curves. Note that for $D = 2, 3$ and $g = 1$, the two inflection points lie on the different sides of the $\alpha = t/\tau$ line although both are *close* to it. Compare with Table 3.

| D/g | 1 | 2 |
|---|---|---|
| 1 | - | - |
| 2 | {0.329 | 1/3} | {0.260 | 0.200} |
| 3 | {0.417| 0.4} | {0.365 | 0.250} |

**II. Model verification**

It is rather natural to start the model verification by using the JMAKn model since it is widespread in the crystal growth and glass community [20,44], even wider [21,45] than supposed by its prerequisites. Throughout the remaining part we will use the model with the following expression (compare with [46]):

$$\alpha = 1 - \exp\left[-\left(\frac{2t}{\tau_{JMAKn}}\right)^n\right] \qquad (26)$$

The time is multiplied here by the factor of 2 thus bringing the inflection points of the expression close to the line $\alpha = t/\tau_{JMAK}$, see Figure 3.

## II.1 The differential form of JMAKn

In this subsection we study the differential form of the JMAKn model with the primary goal to find the correspondence between the two time scales - $\tau_{Dg}$ and $\tau_{JMAKn}$. First we find an expression for $t(\alpha)$:

$$t/\tau_{JMAK} = \frac{1}{2}[-\ln(1-\alpha)]^{1/n} \qquad (27)$$

in order to differentiate it with respect to $\alpha$:

$$\frac{d(t/\tau_{JMAK})}{d\alpha} = \frac{[-\ln(1-\alpha)]^{\frac{1-n}{n}}}{2n(1-\alpha)} \qquad (28)$$

and then to find the time derivative of $\alpha$ according to *the inverse function theorem*:

$$\frac{d\alpha}{d(t/\tau_{JMAK})} = 2n(1-\alpha)\left[\ln\left(\frac{1}{1-\alpha}\right)\right]^{\frac{n-1}{n}} \qquad (29)$$

For a similar expression in 2D see [47], also [19] and their eq.18, but formulated in terms of a probability for given $\alpha$.

Note that $\frac{1}{1-\alpha} > 1$ when $0 < \alpha < 1$.

Now we expand in Taylor series the logarithm in (29) about $\alpha = 0$:

$$\ln\left(\frac{1}{1-\alpha}\right) = \sum_{k=1}^{\infty} \frac{\alpha^k}{k}, \ |\alpha| < 1 \tag{30}$$

Truncating (30) only to the linear in $\alpha$ term, i.e. $k = 1$, we obtain:

$$\frac{d\alpha}{d(t/\tau_{JMAKn})} = 2n\alpha^{\frac{n-1}{n}}(1-\alpha) \tag{31}$$

to recover with $g = 1$ and $D = n$ and, in particular, to get:

$$\tau_{JMAKn} \equiv \tau_{D1}$$

One should keep in mind that the above result holds when $\alpha \to 0$ and we will see below that the two time scales differ overall by a factor of around 1.1 for $D = 2, 3$ and $g = 1$:

$$\tau_{JMAKn} \approx 1.1\tau_{D1} \tag{32}$$

## II. 2 Inflection points of JMAKn

The position of the inflection point is of primary importance when deciding to fit data with a model that possesses an inflection point but the data still does not. Also, the different positions of the inflection points of two models that are to be compared (cross-fitted) point at difference in their time scales and one could judge this difference directly. The time to achieve the inflection point could be used for non-dimensionalization of a model, logistic [48] or JMAKn [33].

It is straightforward to derive the inflection points $\{(t/\tau_{JMAKn})*|\alpha*\}$ of JMAKn as:

$$\{(t/\tau_{JMAKn})*|\alpha*\} = \left\{\frac{1}{2}\left(\frac{n-1}{n}\right)^{\frac{1}{n}} | 1-e^{\frac{1-n}{n}}\right\} \tag{33}$$

In the form of an expression, eq. is presented as:

$$t*(\alpha*) = \frac{1}{2}\left(\log\left(\frac{1}{1-\alpha*}\right)\right)^{1+\log(1-\alpha*)}, \ 0 < \alpha^* \leq 1 - \frac{1}{e} \tag{34}$$

See in Table 3 these for some chosen values of $n$ and comparison with the inflection points of $\alpha_{Dg}$ in Figure 3. This is the small difference in the numerical values of the inflection points that is to be compensated when fitting one of the models with the other leading to the observation .

Table 3. Inflection points of the JMAKn model. The comparison with the inflection points of $\alpha_{Dg}$ is illustrated in Figure 3.

| n | 1 | 1.725 | 2 | 2.43 | 3 | 4 |
|---|---|---|---|---|---|---|
| $\{(t/\tau_{JMAKn})*|a*\}$ | - | {0.303\|0.343} | {0.354\|0.393} | {0.402\|0.445} | {0.437\|0.487} | {0.465\|0.528} |

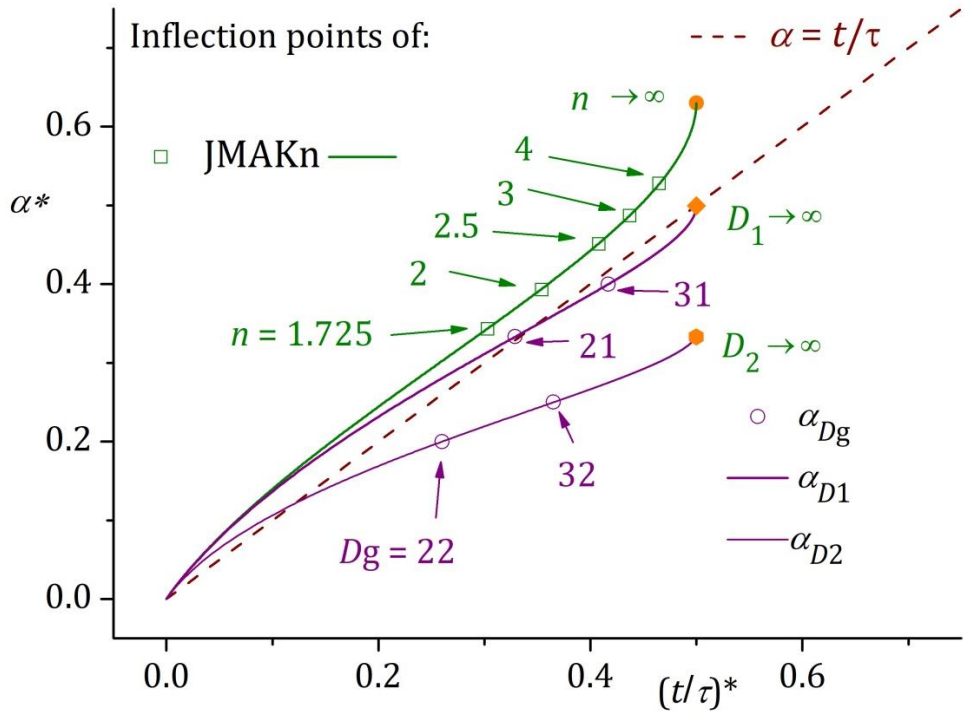

Figure 3. Inflection points of the two models – JMAKn, eq. and $\alpha_{Dg}$, eq..

## II.3 Verification: fitting $\alpha_{Dg}$ with JMAKn

In this subsection we will fit the two most interesting realizations of $\alpha_{Dg}$ - $\alpha_{21}$ and $\alpha_{31}$ with the JMAKn model in the form of . For the former we have (see Table 1) an analytic expression in terms of $\alpha_{21} = \tanh^2 2t/\tau_{21}$ while for the latter we have only the dependence $t = \tau_{31} f(\alpha_{31})$. Important aspect of the fitting between models is that when non-dimensional expressions are used, the scales used for non-dimensionalization, in our case the time-scales, are not necessarily the same. Therefore, (dimensionless) conversion factors must be used:

$$\tau_{JMAKn} = c_f \tau_{Dg} \tag{35}$$

and then the fitting function that uses JMAKn becomes:

$$\alpha = 1 - \exp\left[-\left(\frac{2t/\tau_{Dg}}{c_f}\right)^n\right] \tag{36}$$

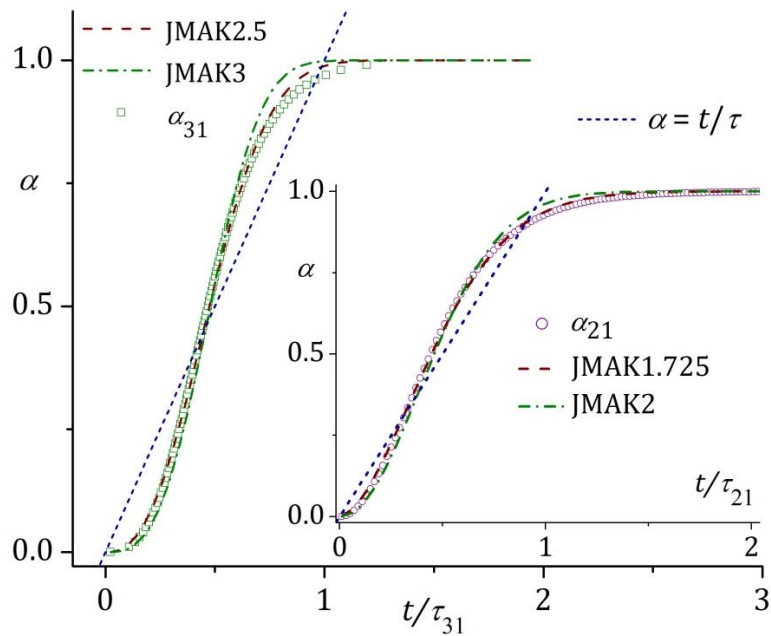

Figure 4. Fitting $\alpha_{21}$ and $\alpha_{31}$ with JMAKn, in both cases the time scale $\tau_{JMAKn} = c_f \tau_{Dg}$, with $c_f \approx 1.1$.

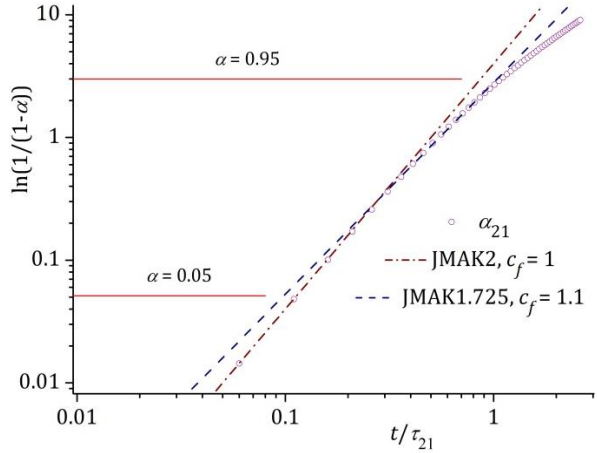
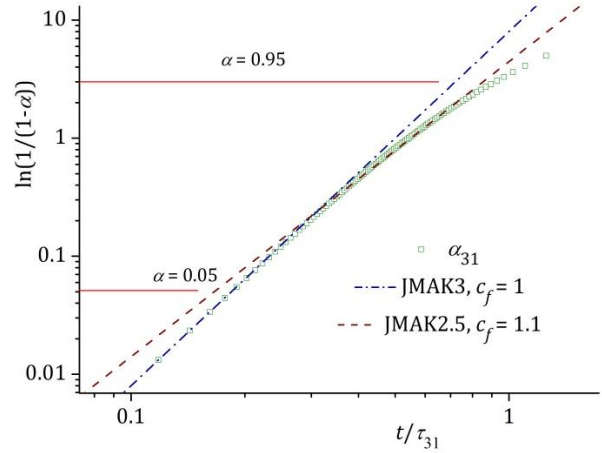

Figure 5. Avrami plot of $\alpha_{21}$ together with JMAK with two exponents $n$=1.725 and 2, note the different conversion factors in the two cases – while JMAK1.725 fits the overall curve, JMAK2 is drawn with $c_f$ = 1.

Figure 6. Avrami plot of $\alpha_{31}$ together with JMAK with two exponents $n$=2.5 and 3, note the different conversion factors in the two cases but same as in Figure 7. Only when $\alpha$ is close to 0, $D=n$.

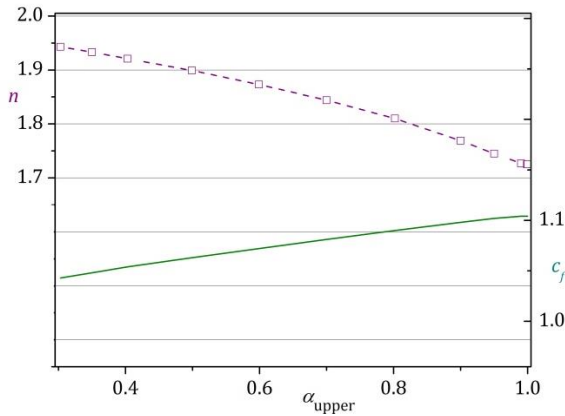
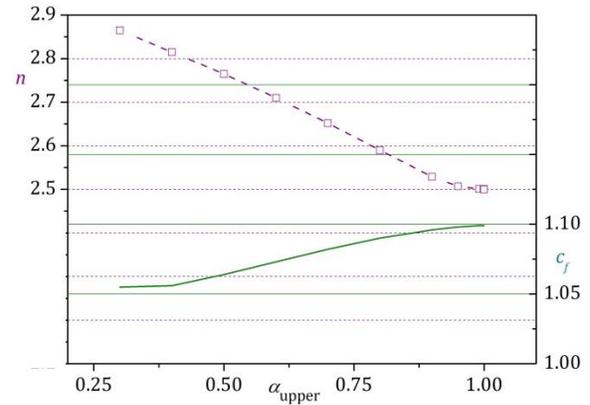

Figure 7. Changing the upper threshold of the fitting interval of $\alpha_{21}$ with JMAKn, from $\alpha_{upper} = 0.3$ up to $\alpha_{upper} = 0.999$ changes the found values of $n$, left y-axis, and the conversion factor, lower dependence, right y-axis obtained. In all cases the starting value of $\alpha_{21} = 0$.

Figure 8. Changing the upper threshold of the fitting interval of $\alpha_{31}$ with JMAKn, from $\alpha_{upper} = 0.3$ up to $\alpha_{upper} = 0.999$ changes the values of $n$, upper dependence, left y-axis, and the conversion factor, lower dependence, right y-axis obtained. In all cases the starting value of $\alpha_{31} = 0$.

Note the use of the so called Avrami plot [22] technique, Figure 5 and Figure 6, to distinguish between the models.

The importance of the upper value of $\alpha$ to which is fitted a dataset was realized in the past [50] but is still not part of the fitting protocols. Here this is quantified in Figure 7 and

Figure 8. In all the fits the initial value of $\alpha$ is 0. These two figures contain a clear message – when fitting data obtained from experiments the results depend on the range of data and they could mask the true model that applies in the concrete case. Here it should be noted that the "globally optimal" ($\alpha_{upper} \approx 1$) set of fitting parameters ($D$, $g$, $c.f. \approx 1.1$) for a particular value of *n* in JMAKn, lead to a $\alpha_{Dg}$ curve that is metrically (the so called 'l2 – norm') close to the curve of JMAKn, but it does not interpolate it at $t = 0$. Conversely, the Taylor* expansion - $D = n$, $g = 1$, $c_f = 1$, leads to a curve that is very close to JMAKn at the beginning of the process (interpolates JMAKn at 0) and diverges away from it for larger transformation ratios.

## III. Towards model validation

The validation of a new model is not a singular event but a process that takes many stages and our aim here is rather to build the lines of validation towards a unified protocol of validation that will permit to place the further studies in a ready frame.

### III. 1. Fitting back JMAKn with $\alpha_{Dg}$

The validation of a model is a matter of a longer development than the time scale of a single paper. Here we use the developed numerical procedure (see Appendix 1) in which the parameters of our model – $D$, $g$ and $\tau_{Dg}$, are allowed to be real-valued and thus to serve as fitting parameters. We are going to fit datasets prepared from the JMAKn model with chosen values of *n* and with $\tau_{JMAKn} = 1$. Throughout this session we will fix $g = 1$ since we already have seen, eq.(29), that JMAKn contains also the $(1-\alpha)$ term. This is, in fact, an important simplification before relaxing *g* too since we do not know at all the numerical behavior of our procedure and, thus, of the model. The first datasets prepared are with an "ideal" interval of the transformation ratio values - $\alpha \in [0, 0.9999]$.

Table 4. Values for *D* and the conversion factor $c_f$ resulting in from fitting datasets obtained using JMAKn with typical values of *n*, with values of $\alpha \in [0, 0.9999]$, *g* is fixed to 1 throughout the fitting session, $\tau_{JMAKn} = 1$. The datasets with *n*=3, 4 are also fitted but only to illustrate the divergence between the two models with the departure from *n*=*D*=1.

| *n* from JMAKn | *D* | $c_f$ | $R^2$ | Procedure |
|---|---|---|---|---|
| 1 | 1.002 | 1.00026 | 0.9999992 | NLSQ |
|   | 1 | 0.9999 | 0.99999998 | UNIFORM |
| 1.725 | 1.989 | 1.104 | 0.9994 | NLSQ |
|   | 1.966 | 1.105 | 0.9993 | UNIFORM |
| 2 | 2.371 | 1.109 | 0.9991 | NLSQ |
|   | 2.352 | 1.110 | 0.9991 | UNIFORM |
| 2.43 | 2.993 | 1.105 | 0.99883 | NLSQ |
|   | 3.000 | 1.108 | 0.99882 | UNIFORM |
| 2.5 | 3.073 | 1.106 | 0.99879 | NLSQ |
|   | 3.036 | 1.107 | 0.99884 | UNIFORM |
| 3 | 3.794 | 1.100 | 0.99859 | NLSQ |
|   | 3.801 | 1.102 | 0.99857 | UNIFORM |
| 4 | 5.215 | 1.083 | 0.99828 | NLSQ |
|   | 5.267 | 1.084 | 0.99829 | UNIFORM |

### III. 2. Validation in 2D

In 2005 Min et al.[21] used TEM to study the kinetics of crystallization in a quasi-2D system – ALD $Ta_2O_5$ films deposited on Si substrates. Using the difference between the crystal and amorphous phase in the TEM images they found for three different selected temperatures – 790°C, 820°C and 850°C, different values of *n* = 2.5, 1.9 and 1.7, see Figure 9. It is clearly seen that the rescaled data in the original data ranges do not collapse on a single master curve due to the differences in *n*. We digitized their data, fig.7 in [26], and fitted it with $\alpha_{21}$ as shown in Figure 10.

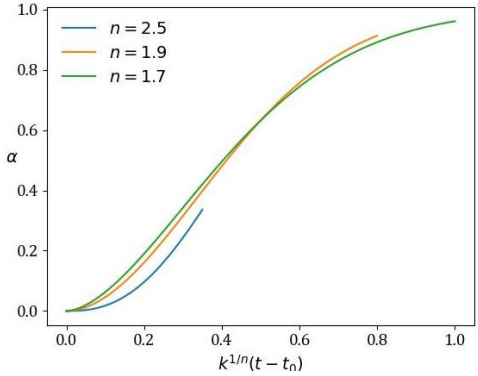 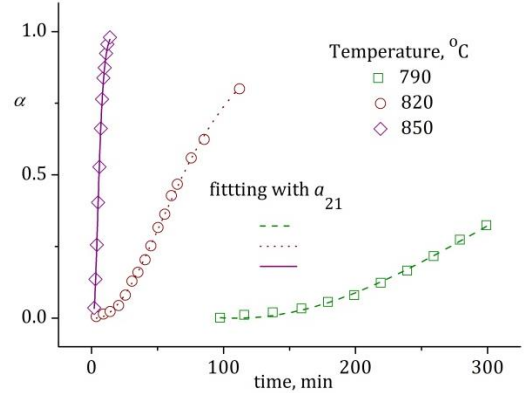

Figure 9. The three values for *n* found in [21] when fitting the crystallization data with JMAKn are used here to illustrate how would look the rescaled data, with $\alpha_{upper}$ = ~0.4 (790°C), ~0.8 (820°C), ~1 (850°C)

Figure 10. The crystallization data from [26] fitted with $\alpha_{21}$, $R^2$ = 0.996, 0.999, 0.999, respectively. The scale $\tau_{21}$ and the induction time $t_0$ are not given since there is no theory to be evoked for their explanation.

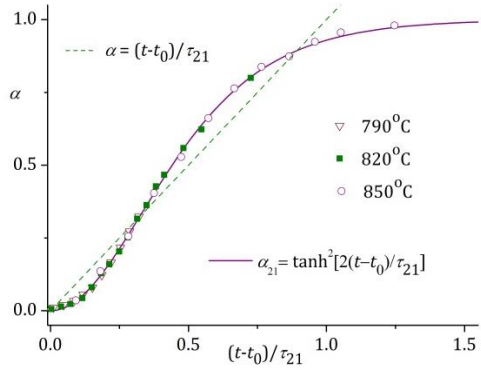 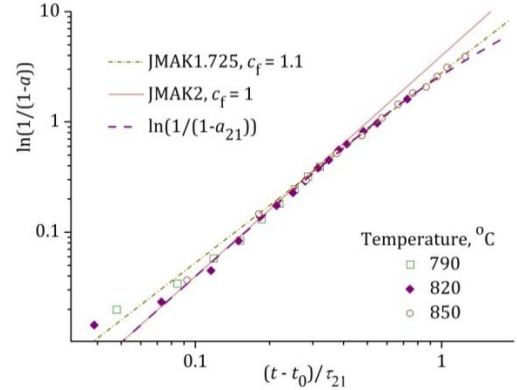

Figure 11. Crystallization data from [26] rescaled with the fit parameters from $\alpha_{21}$.

Figure 12. Rescaled data from Figure 11 plotted in Avrami-coordinates. It is clearly seen that the data follows the general direction of $\alpha_{21}$.

More data sets are successfully analyzed using the same protocol, for example [51,52], but the results will be published elsewhere.

## III. 3. Towards a general validation protocol

The protocol of model validation should probably start with a shortcut to the procedure described above and using the "canonical" curves in Figure 7 and Figure 8 as illustrated in Figure 13. Such a preliminary check does not require additional tools but only a simple procedure of multiple fitting the experimental data with JMAKn changing the upper value of the fitting interval $\alpha_{upper}$.

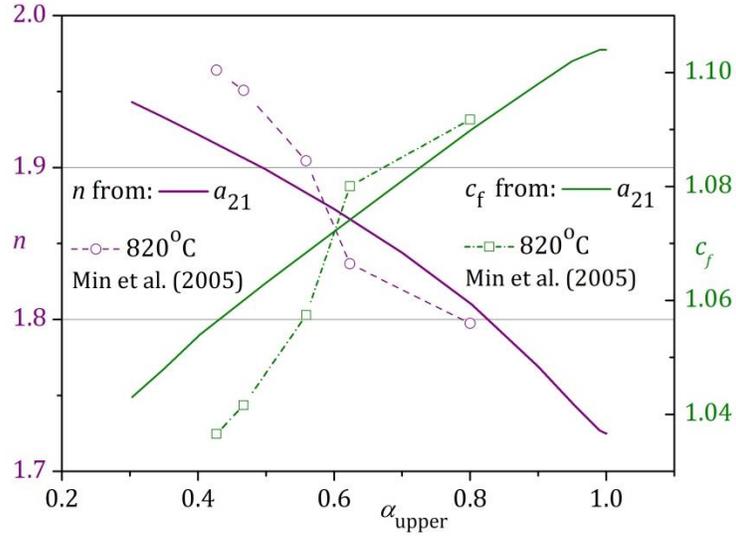

Figure 13. Comparison between the predicted behavior of Avrami exponent $n$ and the conversion factor $c_f$ between the two timescales of $\alpha_{21}$ and JMAKn and the behavior found from the crystallization data reported in [26]. In order to obtain $c_f$ from the experimental data the value of $\tau_{21}$ found from fitting the whole dataset, up to $\alpha=0.8$, is used.

Apart from this easy check, one should judge whether the data is collected from a system that corresponds to the prerequisites of our model – D>1, rate of transformation that depends on the supersaturation, diffusion-limited regime of growth ( $g=1$ ) throughout the whole process. Then, one takes the experimental dataset and fits it with JMAKn. If the transformation is almost completed, $\alpha \rightarrow 1$ the values of $n$ expected are $n \rightarrow 2.4-2.5$ in 3D and $n \rightarrow 1.725$ in 2D. Still, $n$ below 3 is within the range of our model and experiments done in 3D are to be preferred since there the difference between the expected $n=3$ and the obtained one is bigger. Then, one fits again the data with both JMAKn and $\alpha_{D1}$ but this time fixing $n$ and $D$ to obtain the time scales from the two models, the values of $R^2$ could serve for initial judgment. Then one rescales the time of the dataset separately by the obtained time scales and plots in separate plots the result for JMAKn and $\alpha_{D1}$ together with the curves of the corresponding models to see how the data points collapse on them. Yet at that stage one should be able to judge which of the two models performs better. As a last tool, one can plot the two corresponding rescaled datasets in Avrami [22] plots – the data that are described better by JMAKn remain ordered along a straight line there while the data that fits better with $\alpha_{D1}$ is "turning" clock-wise, starting from the corresponding *n* and turns almost gradually towards lower values of *n* achieving at the end even lower ones compared with the value of *n* that best fits the $\alpha_{D1}$ model overall, 1.725 of 2.5 in the corresponding dimension.

**IV. Discussion and Conclusion**

In this paper we derive from first principles a model to model the crystal growth dynamics in conditions in which the supersaturation is raised only at the beginning of the process and is not sustained further, so, it expires due to the crystallization. After formulating the differential equation that describes the process (17) for chosen combination of spatial dimension and growth order we integrate it we derive expressions for the time evolution of the transformation ratio, Table 1. Further we study our model in parallel with the model of Johnson-Mehl-Avrami-Kolmogorov and one of the practical reasons to use JMAKn to fit our expressions for $\alpha_{D1}$, D =2, 3, see Figure 4, is to bracket to range of candidates to be studied by our model, and more specifically those that are resulting in non-integer values of *n* when fitted with JMAKn. We also show, by obtaining the differential form of JMAKn and expanding it in Taylor series around $\alpha$=0 that $\alpha_{D1}$, D =2, 3 coincide with *n*=2, 3 including identity of the two timescales $\tau_{D1} = \tau_{JMAKn}$.

Fitting successfully published experimental data with $\alpha_{21}$ we lay down the directions of building a general protocol for validating and using further our model.

An interesting corollary of our investigation (and the recipe suggested in III.3) is that the data points that have the largest discriminatory power for the different models are those close to the end of the process ($\alpha \to 1$). This can be seen both analytically (JMAKn's Taylor series vs. $\alpha_{D1}$) or numerically (the "turn" in Avrami plots). Such a conclusion is somewhat unintuitive – one expects that close to the equilibrium all differences should "even out", while the quick power law growth ($\alpha \sim t^D$) at the beginning should be where differences between growth regimes should be most pronounced.

Thus, the widespread usage of JMAKn could be rationalized on the basis of this conclusion. Crystallization is rarely driven to completion, especially when the timescale is on the order of days. This, combined with the relaxation on the requirement for the exponent *n* to be an integer value has made the JMAKn to produce "artificially good" numeric results, while the understanding of the growth process has been somewhat left behind. This should be taken as a general principle when modelling sigmoid growth – while a lot of sigmoid curves would most likely yield good fitting results, without keeping in mind the assumptions under which they were derived, one might find themselves attempting to "fit" the experiment to the model, instead of the opposite.

**Acknowledgements.** VT acknowledges the Mercator Fellowship from the German Research Foundation (DFG) at the Institute for Materials Research and Testing (BAM), and the Glass Department at BAM for the warm hospitality during his stay in Berlin. VT also thanks Christo Nanev (Sofia) for introducing him to the exciting field of protein nucleation and crystallization, to Lyuba Dimova, Julia Romanova and Hristina Popova (Sofia) and Ralf Mueller (Berlin) for the interest in this study. VVI acknowledges the COST-CA18234 for the opportunity to take part in the "COST Training School on Computational Materials Modeling". Part of the calculations was done on the HPC facility Nestum (BG161PO003-1.2.05).

**Dedication**

This paper is dedicated to the memory of Isak Avramov (1946–2020).

..

**Appendix 1. Numerical procedures of fitting with $\alpha_{Dg}$**

**Numerical integration**

As it was discussed in the main text of the paper, directly integrating and obtaining a closed-form expression for the curve is not possible in general. To sidestep this problem, we develop a simple numerical procedure for solving the main differential equation of our model, eq. (17), based on Dormand & Prince RK8 (5,3) – DOP853 in Python. In the SciPy library [27] an implementation of DOP853 with dense output can be found. This allows us to run the numeric integration for a given set of parameters - $D, g, \tau_{Dg}$ and time interval - $[t_{initial}, t_{final}]$ once and use 7$^{th}$ – degree polynomial interpolation after that to calculate the value of $\alpha$ for arbitrary values of $t \in [t_{initial}, t_{final}]$.

Using such high order initial-value problem solvers is justified for two reasons – the right-hand side of the only differential equation is "computationally cheap" to evaluate, so using a high order RK method would not impact the time-performance much. On the other hand, this allows us to obtain highly accurate solutions that would make subsequent numerical methods (such as optimizers, equation solvers, etc.) more stable.

All our numerical code is freely available on GitHub. [43]

**Non-linear least squares fit (NLSQ)**

Being able to calculate arbitrary values of $\alpha$ for a given integration interval allows us to proceed with fitting the model parameters $D, g, \tau_{Dg}$ to a given dataset – either experimental data or datapoints generated from another model such as JMAK-n. Here it is important to constrain the optimization problem with the proper bounds for the parameters. A well-suited procedure for such a constrained non-linear least-squares problem is the Trust Region Reflective algorithm (TRF) which is generally robust even when the initial guess is far from the minimum. Again, an implementation of TRF can be found in SciPy which only requires as input the initial guess and a function that calculates the residuals vector for a given parameter set.

This combined procedure can be found in our "*parameter_finder.py*" script, which has been the main numerical core for the present investigation.

**Uniform approximation (UNIFORM)**

We can also rewrite the fitting procedure described above in term of uniform (min-max) approximations. This can be done by directly minimizing the infinity-norm of the residuals vector (instead of the Euclidean norm) using the Nelder-Mead simplex algorithm provided in SciPy. This can be useful, since a least-squares fit guarantees that "the average error" is small, while an uniform approximation guarantees that at all points the error is bounded by some maximum value. Even though the maximum norm is continuous, it is not smooth. This has the side effect that the solution might not be unique, and it can pose serious problems for a gradient-based optimizer. That is why a simplex algorithm has been used. An implementation of the uniform approximation method can be found in the "*parameter_finder_uniform.py*" script.

[1]     A. Chernov, *Formation of Crystals in Solutions*, Contemporary Physics **30**, 251 (1989).

[2]     W. Kossel, *On the Theory of Crystal Growth*, News from the Göttingen Society of Sciences, Mathematical-Physical Class **1927**, 135 (1927).

[3]     I. N. Stranski, *Zur Theorie Des Kristallwachstums*, Zeitschrift Für Physikalische Chemie **136**, 259 (1928).

[4]     R. Kaischew, *On the History of the Creation of the Molecular-Kinetic Theory of Crystal Growth: Honoring the Memory of IN Stranski*, (1981).

[5]     V. L. Tassev and D. F. Bliss, *Stranski, Krastanov, and Kaischew, and Their Influence on the Founding of Crystal Growth Theory*, Journal of Crystal Growth **310**, 4209 (2008).

[6]     T. Yamamoto, Y. Akutsu, and N. Akutsu, *Universal Behavior of the Equilibrium Crystal Shape near the Facet Edge. I. A Generalized Terrace-Step-Kink Model*, Journal of the Physical Society of Japan **57**, 453 (1988).

[7]     W.-K. Burton, N. Cabrera, and F. C. Frank, *The Growth of Crystals and the Equilibrium Structure of Their Surfaces*, Philosophical Transactions of the Royal Society of London A: Mathematical, Physical and Engineering Sciences **243**, 299 (1951).

[8]     A. A. Chernov, *Notes on Interface Growth Kinetics 50 Years after Burton, Cabrera and Frank*, Journal of Crystal Growth **264**, 499 (2004).

[9]     F. Krzyżewski, M. Załuska-Kotur, A. Krasteva, H. Popova, and V. Tonchev, *Scaling and Dynamic Stability of Model Vicinal Surfaces*, Crystal Growth & Design **19**, 821 (2019).

[10]    M. Za\luska-Kotur, H. Popova, and V. Tonchev, *Step Bunches, Nanowires and Other Vicinal "Creatures"—Ehrlich–Schwoebel Effect by Cellular Automata*, Crystals **11**, 1135 (2021).

[11]    I. Stranski and R. Kaischew, *Gleichgewichtsformen Homöopolarer Kristalle*, Zeitschrift Für Kristallographie-Crystalline Materials **78**, 373 (1931).

[12]    T. A. Witten Jr and L. M. Sander, *Diffusion-Limited Aggregation, a Kinetic Critical Phenomenon*, Physical Review Letters **47**, 1400 (1981).

[13]    B. Ranguelov, D. Goranova, V. Tonchev, and R. Yakimova, *Diffusion Limited Aggregation with Modified Local Rules*, Comptes Rendus de l'Academie Bulgare Des Sciences/Proceedings of the Bulgarian Academy of Sciences **65**, 913 (2012).

[14]    A. N. Kolmogorov, *On the Statistical Theory of the Crystallization of Metals*, Bull. Acad. Sci. USSR, Math. Ser **1**, 355 (1937).

[15]    W. Johnson and R. Mehl, *Trans*, in *AIME*, Vol. 135 (1939), p. 416.

[16]    M. Avrami, *Kinetics of Phase Change. I General Theory*, The Journal of Chemical Physics **7**, 1103 (1939).

[17]    J. Christian, *The Theory of Transformations in Metals and Alloys* (Newnes, 2002).

[18]    N. V. Alekseechkin, *Extension of the Kolmogorov–Johnson–Mehl–Avrami Theory to Growth Laws of Diffusion Type*, Journal of Non-Crystalline Solids **357**, 3159 (2011).

[19]    M. Fanfoni and M. Tomellini, *The Johnson-Mehl-Avrami-Kohnogorov Model: A Brief Review*, Il Nuovo Cimento D **20**, 1171 (1998).

[20]     R. Müller and S. Reinsch, *Viscous Phase Silicate Processing*, in *Processing Approaches for Ceramics and Composites*, Vol. 3 (John Wiley & Sons Hoboken, New Jersey, USA, 2012), pp. 75–144.

[21]     R. Svoboda, *Crystallization of Glasses–When to Use the Johnson-Mehl-Avrami Kinetics?*, Journal of the European Ceramic Society **41**, 7862 (2021).

[22]     A. T. Lorenzo, M. L. Arnal, J. Albuerne, and A. J. Müller, *DSC Isothermal Polymer Crystallization Kinetics Measurements and the Use of the Avrami Equation to Fit the Data: Guidelines to Avoid Common Problems*, Polymer Testing **26**, 222 (2007).

[23]     P. Bruna, D. Crespo, R. González-Cinca, and E. Pineda, *On the Validity of Avrami Formalism in Primary Crystallization*, Journal of Applied Physics **100**, 054907 (2006).

[24]     M. Tomellini, *Mean Field Rate Equation for Diffusion-Controlled Growth in Binary Alloys*, Journal of Alloys and Compounds **348**, 189 (2003).

[25]     A. Burbelko, E. Fraś, and W. Kapturkiewicz, *About Kolmogorov's Statistical Theory of Phase Transformation*, Materials Science and Engineering: A **413**, 429 (2005).

[26]     K.-H. Min, R. Sinclair, I.-S. Park, S.-T. Kim, and U.-I. Chung, *Crystallization Behaviour of ALD-Ta2O5 Thin Films: The Application of in-Situ TEM*, Philosophical Magazine **85**, 2049 (2005).

[27]     J. Málek, *Kinetic Analysis of Crystallization Processes in Amorphous Materials*, Thermochimica Acta **355**, 239 (2000).

[28]     P.-F. Verhulst, *Notice Sur La Loi Que La Population Suit Dans Son Accroissement*, Corresp. Math. Phys. **10**, 113 (1838).

[29]     A. Tsoularis and J. Wallace, *Analysis of Logistic Growth Models*, Mathematical Biosciences **179**, 21 (2002).

[30]     C. N. Nanev, V. D. Tonchev, and F. V. Hodzhaoglu, *Protocol for Growing Insulin Crystals of Uniform Size*, Journal of Crystal Growth **375**, 10 (2013).

[31]     V. Tonchev and C. Nanev, *Growth and Dissolution of Equally-Sized Insulin Crystals*, Crystal Research and Technology **48**, 1003 (2013).

[32]     V. Dubrovskii, *Private Communication*, (unpublished).

[33]     D. Kashchiev, *Kinetics of Protein Fibrillation Controlled by Fibril Elongation*, Proteins: Structure, Function, and Bioinformatics **82**, 2229 (2014).

[34]     A. A. Chernov, L. N. Rashkovich, and A. A. Mkrtchan, *Solution Growth Kinetics and Mechanism: Prismatic Face of ADP*, Journal of Crystal Growth **74**, 101 (1986).

[35]     S. Stoyanov and V. Tonchev, *Properties and Dynamic Interaction of Step Density Waves at a Crystal Surface during Electromigration Affected Sublimation*, Physical Review B - Condensed Matter and Materials Physics **58**, 1590 (1998).

[36] V. Tonchev, B. Ranguelov, H. Omi, and A. Pimpinelli, *Scaling and Universality in Models of Step Bunching: The "c +-C -" Model*, European Physical Journal B **73**, 539 (2010).

[37] A. Jouyban, *Handbook of Solubility Data for Pharmaceuticals* (CRC press, 2009).

[38] E. Meron, *Pattern Formation in Excitable Media*, Physics Reports **218**, 1 (1992).

[39] G. A. Hirata, A. Bernardo, and E. A. Miranda, *Determination of Crystal Growth Rate for Porcine Insulin Crystallization with CO2 as a Volatile Acidifying Agent*, Chemical Engineering and Processing: Process Intensification **56**, 29 (2012).

[40] G. I. Barenblatt, *Scaling, Self-Similarity, and Intermediate Asymptotics: Dimensional Analysis and Intermediate Asymptotics*, Vol. 14 (Cambridge University Press, 1996).

[41] I. Avramov, K. Avramova, and C. Rüssel, *New Method to Analyze Data on Overall Crystallization Kinetics*, Journal of Crystal Growth **285**, 394 (2005).

[42] J. Šesták and I. Avramov, *Rationale and Myth of Thermoanalytical Kinetic Patterns: How to Model Reaction Mechanisms by the Euclidean and Fractal Geometry and by Logistic Approach*, in *Thermal Physics and Thermal Analysis* (Springer, 2017), pp. 295–318.

[43] V. Ivanov, *Vasilvas99/Sigmoid-Tools: Sigmoid Tools Archive 0.1*, (2022).

[44] J. Amorós, E. Blasco, A. Moreno, N. Marín, and C. Feliu, *Sinter-Crystallisation Kinetics of a SiO2–Al2O3–CaO–MgO–SrO Glass-Ceramic Glaze*, Journal of Non-Crystalline Solids **532**, 119900 (2020).

[45] I. Avramov, *Kinetics of Distribution of Infections in Networks*, Physica A: Statistical Mechanics and Its Applications **379**, 615 (2007).

[46] I. Avramov, K. Avramova, and C. Rüssel, *New Method to Analyze Data on Overall Crystallization Kinetics*, Journal of Crystal Growth **285**, 394 (2005).

[47] S. Stoyanov, *Layer Growth of Epitaxial Films and Superlattices*, Surface Science **199**, 226 (1988).

[48] C. N. Nanev and V. D. Tonchev, *Sigmoid Kinetics of Protein Crystal Nucleation*, Journal of Crystal Growth **427**, 48 (2015).

[49] M. Lambrigger, *Avrami Master Curves for Isothermal Polymer Crystallization*, Polymer Journal **29**, 188 (1997).

[50] T. Pradell, D. Crespo, N. Clavaguera, and M. Clavaguera-Mora, *Diffusion Controlled Grain Growth in Primary Crystallization: Avrami Exponents Revisited*, Journal of Physics: Condensed Matter **10**, 3833 (1998).

[51]     D. Tatchev, T. Vassilev, G. Goerigk, S. Armyanov, and R. Kranold, *Kinetics of Primary Crystallization of Hypoeutectic Amorphous Ni–P Alloy Studied by in Situ ASAXS and DSC*, Journal of Non-Crystalline Solids **356**, 351 (2010).

[52]     H. Choi, Y. Kim, Y. Rim, and Y. Yang, *Crystallization Kinetics of Lithium Niobate Glass: Determination of the Johnson–Mehl–Avrami–Kolmogorov Parameters*, Physical Chemistry Chemical Physics **15**, 9940 (2013).

[53]     P. Virtanen et al., *SciPy 1.0: Fundamental Algorithms for Scientific Computing in Python*, Nature Methods **17**, 261 (2020).